\documentclass[twocolumn,twoside]{IEEEtran}

\usepackage{setspace}

\usepackage{flushend}

\usepackage{mathpazo}
\usepackage{times}
\usepackage{amsmath}
\usepackage{amsfonts}
\usepackage{latexsym}
\usepackage{amssymb}
\usepackage{cite}

\usepackage{upref}
\usepackage{theorem}
\usepackage{graphicx}
\usepackage{psfrag}
\usepackage{color}

\usepackage[inline]{trackchanges}
\usepackage{setspace}

\newcommand\isitremove[1]{#1}

\addeditor{alex}
\addeditor{}

\newtheorem{alg}{Algorithm}

\newtheorem{theorem}{Theorem}
\newtheorem{lem}{Lemma}
\newtheorem{clm}[lem]{Claim}
\newtheorem{definition}{Definition}

\newtheorem{coro}{Corollary}

\newcommand{\cA}{{\cal A}}

\newcommand{\cD}{{\cal D}}

\newcommand{\BA}{\begin{alg}} \newcommand{\EA}{\end{alg}}
\newcommand{\BE}{\begin{enumerate}} \newcommand{\EE}{\end{enumerate}}
\newcommand{\BT}{\begin{theorem}} \newcommand{\ET}{\end{theorem}}
\newcommand{\BL}{\begin{lem}} \newcommand{\EL}{\end{lem}}
\newcommand{\BCM}{\begin{clm}} \newcommand{\ECM}{\end{clm}}
\newcommand{\BCR}{\begin{coro}} \newcommand{\ECR}{\end{coro}}

\newcommand{\BI}{\begin{itemize}} \newcommand{\EI}{\end{itemize}}

\def\FullBox{\hbox{\vrule width 8pt height 8pt depth 0pt}}
\newcommand{\qed}{\;\;\;\FullBox}
\newenvironment{prf}{\noindent{\bf Proof:~~}}{\(\qed\)}
\newcommand{\BPF}{\begin{prf}} \newcommand {\EPF}{\end{prf}}
\newenvironment{proofof}[1]{\noindent{\bf Proof of {#1}.~}}{\endprf}
\newcommand{\BPFOF}{\begin{proofof}} \newcommand {\EPFOF}{\end{proofof}}

\newcommand{\eat}[1]{}
\newcommand{\eps}{\varepsilon}

\usepackage{stackengine}

\begin{document}

\title{{On Subset Retrieval and Group Testing Problems with Differential Privacy Constraints}}


\author{\large
Mira Gonen\thanks{Mira Gonen is with the School of Computer Science, Ariel University, Ariel, 40700, Israel (e-mail: mirag@ariel.ac.il).},
Michael Langberg\thanks{Michael Langberg is with the Department
  of Electrical Engineering, State University of New-York at Buffalo, Buffalo, NY 14260, USA (e-mail: mikel@buffalo.edu). Work supported in part by NSF grant 2245204.}, and
Alex Sprintson\thanks{Alex Sprintson is with the Department of Electrical and Computer Engineering, Texas A\&M University, College Station, TX 77843-3128, USA (e-mail:  alex.sprintson@tamu.edu). }}
\maketitle

\thispagestyle{empty}

\begin{abstract}
This paper focuses on the design and analysis of privacy-preserving techniques for group testing and infection status retrieval. Our work is motivated by the need to provide accurate information on the status of disease spread among a group of individuals while protecting the privacy of the infection status of any single individual involved.  The paper is motivated by practical scenarios, such as controlling the spread of infectious diseases, where individuals might be reluctant to participate in testing if their outcomes {are not kept confidential}.

The paper makes the following contributions. First, we present a differential privacy framework for the \emph{subset retrieval problem}, which focuses on sharing the infection status of individuals with administrators and decision-makers. We characterize the trade-off between the accuracy of 
subset retrieval
and the degree of privacy guaranteed to the individuals. In particular, we establish tight lower and upper bounds on the achievable level of accuracy subject to the differential privacy constraints.  We then formulate the {differential} privacy framework for 
 the noisy group testing problem in which noise is added either before or after the pooling process.
We establish a reduction between the private subset retrieval and noisy group testing problems and show that the converse and achievability schemes for subset retrieval  carry over to  {differentially} private group testing. 
\end{abstract}

\section{Introduction}

Since the recent pandemic, there has been a need to develop methods and tools to ensure the privacy of all individuals. This is motivated by the ethical, legal, and regulatory requirements to protect individuals' infection status, susceptibility to infection, social interactions, and prior contact histories. Indeed, numerous laws and regulations, such as HIPAA and GDPR, are in place to protect sensitive information.  Protecting privacy offers numerous benefits, including fostering trust between individuals, their employers, health authorities, and medical professionals (see e.g., \cite{9627089,Bengio2020,morales2021covid19testsgonerogue,Tiffany_Li_2021} and references therein). 

This work focuses on the security and privacy of the testing and infection status disclosure process, which are instrumental in ensuring greater participation in testing and data sharing programs conducted by employers and health agencies. Indeed, many individuals are less likely to 
cooperate in the infection mitigation process
if their infection status becomes public knowledge. 
 %
For example, consider an educational institution, such as a school, where the administration seeks to assess the current status of disease spread. The administration’s objective could be, for example,
identifying the classes and student groups most affected by the spread of the infection. Furthermore, since testing is often conducted voluntarily, many students may hesitate to participate if there is a risk that their results could be disclosed to peers or instructors. Therefore, protecting privacy is essential to ensuring participation and compliance with applicable laws and regulations. 

In this paper, we introduce privacy-preserving frameworks for two key problems: the \emph{subset retrieval problem}, which focuses on sharing the infection status of individuals with administrators and decision-makers, and the \emph{group testing problem}, which aims to assess the spread of disease through pooled testing. 
%
%
 We adopt the framework of \emph{differential privacy}~\cite{10.1007/978-3-540-79228-4_1} to define and analyze a formal set of privacy and accuracy conditions. Our framework effectively balances the trade-off between protecting individual privacy and maintaining sufficient accuracy in the results. Our goals include (i)~characterizing the fundamental limits and the trade-off between privacy and accuracy that can be achieved by a privacy-preserving scheme and (ii) developing a practical privacy-preserving schemes with optimal or near-optimal performance guarantees. 

\textbf{Differentially Private Subset Retrieval (DPSR).} The DPSR problem includes a set $V$ of $n$ individuals and two entities, a \emph{database} that knows the set $e\subset V$ of infected individuals, and an  \emph{agent} that seeks to retrieve $e$. The goal is to find a probabilistic scheme $\cA(e)$ that, for any set $e\in V$, returns a 
set $e'\in V$, such that: 
(i)  $e'$ is {\em close} to $e$ (accuracy) and (ii)  $e'$ hides the status of any single individual (privacy), i.e., it obscures the fact whether any individual $i\in V$ belongs to $e$ or not. The key aspect in the design of such a scheme is to leverage randomness (added noise) to obscure the set $e$. 

 This problem is motivated by the settings in which a medical office in an organization (the 
 database) 
 needs to share the infection status of all individuals with the administration without violating the individuals' privacy. 


\textbf{Differentially-Private Group Testing.} 
Group testing is a fundamental technique for efficiently identifying infected individuals within a given population using the minimum number of pooled tests (see e.g., \cite{10146331,8550650, 9413685}). In this approach, each pooled test is conducted on a subset of individuals and yields a positive result if and only if at least one individual in the group is infected. 



Group testing has a significant potential to guarantee privacy by {avoiding individual tests and, instead, relying on pool testing to estimate the affected groups. In addition, the group testing procedure can take advantage of the inherent noise associated with pooled testing to ensure the privacy of individuals.} 
%
 %
In this work, we present a privacy framework for the group testing problem based on the notion of differential privacy~\cite{10.1007/978-3-540-79228-4_1}.








 \textbf{Contribution.} The paper's contributions can be summarized as follows. First, we establish the fundamental trade-off between accuracy and security for the DPSR problem. In particular, for a given distortion parameter $\beta$, we establish a lower bound on the differential privacy parameter $\delta$ that can be achieved by any scheme. In addition, we present an achievability scheme whose performance is very close to optimal. We  consider an important variation of the DPSR problem (referred to as DPSR-I) in which the database needs to obscure the set $e$  in a way that is independent of $e$, i.e., without relying on the detailed knowledge of $e$.
Finally, we formalize the differential privacy framework for group testing and establish connections between the DPSR, DPSR-I, and private group testing problems. This connection enables us to derive fundamental limits and develop achievable schemes for private group testing.
  %
 
 



\textbf{Related work.}  
There is a substantial body of research that explores various aspects of differential privacy in which database entries are protected in the process of (multiple) functional queries (see, e.g., \cite{10.1007/11681878_14,10.1007/978-3-540-79228-4_1,dwork2014algorithmic,vadhan2017complexity, 10.1145/773153.773173} and references therein). 
%
%
Prior works that discuss both group testing and differential privacy, 
use group testing as a mechanism to achieve differential privacy over the data in various problems. (see, e.g., \cite{bsw17,k13,hkr12}).
Cohen et al.~\cite{ccg21} present techniques to secure group testing against eavesdroppers. Their algorithm is secure in the sense that the eavesdropper cannot learn which of the participants is infected. Atallah et al.~\cite{afbc08} describe a private group {testing} protocol between two entities, Alice and Bob, that informs who is infected without Alice or Bob learning this. 
 However, to the best of our knowledge, this paper is the first to conduct a comprehensive analysis of subset retrieval and group testing within the framework of differential privacy.



\section{Model and Statement of Results}\label{sec:model} 


As mentioned in the Introduction, the DPSR problem  includes two entities, a {\em database} that holds a subset $e \subseteq V$ taken from a ground set of size $n$, and an {\em agent} that knows the ground set $V$ and is interested in retrieving the subset $e$. Subset $e$ may correspond to a collection of individuals with certain (private) attributes. For example, 
the database may hold the identity of a collection of contaminated individuals.
Roughly speaking, a differentially-private solution to the subset-retrieval problem includes a probabilistic database-answer $\cA$ that for every subset $e$ of the database returns a distribution of answers 
$\cA(e)$ to the agent such that, on one hand, $\cA(e)$ is {\em close} to $e$ (accuracy), while on the other hand $\cA(e)$ hides whether any single individual $i$ is in $e$ or not (privacy). We assume that subsets $e$ of $V$ are of a given size $d$,  where $d$ is known to all parties. We denote by $2^V$ all subsets of $V$, and by $E$ all subsets of $V$ of size $d$. For any two subsets $e_i$ and $e_j$ in $2^V$, we define $$d(e_i,e_j)=\max\{|e_i \setminus e_j|, |e_j\setminus e_i|\}.$$
Let $\mu$ be the number of subsets of size $d$, i.e., $\mu={n \choose d}$.
For every subset $e_i$ of size $d$, we define $N_{\alpha}(e_i)$ as the set of all subsets of size $e$ with distance $\alpha$ from $e_i$, i.e., 

$$
N_{\alpha}(e_i)=\{e_j\ |\ d(e_i,e_j)=\alpha\}.
$$

Note that 

$$|N_{\alpha}(e_i)|={d \choose d-\alpha}{n-d\choose \alpha}={d \choose \alpha}{n-d\choose \alpha}.$$

Definition~\ref{def:gen_gt_private_f} presents a formal framework for  DPSR. 

\begin{definition}
\label{def:gen_gt_private_f}
Given a ground set $V$ of size $n$ and parameters $d\leq n,\eps_1,\eps_2,\delta$ and ,$\beta$, an $(\eps_1,\eps_2,\delta,\beta)_{n,d}$ differentially-private solution to the subset-retrieval problem 
is a randomized algorithm 
\mbox{${\cal A}:E\rightarrow 2^V$} that satisfies the following conditions:
 \begin{itemize}
 \item[(i)] (accuracy) for any 
 $e\in E$, it holds that 
 $$\Pr[d({\cal A}(e),e) \le \beta]\ge 1-\varepsilon_1,$$
 \item[(ii)] (privacy) for any $e_i,e_j\in E$ 
 with $d(e_i,e_j)=1$ and any $S \subseteq 2^V$, 
$$\Pr[{\cal A}(e_i)\in S]\le e^{\eps_2}\cdot\Pr[{\cal A}(e_j)\in S]+\delta.$$ 
 \end{itemize}
Here, the probabilities are taken over the randomness of ${\cal A}$.
\end{definition}



Our main results are outlined in Theorem~\ref{the:main-1}.

\BT
\label{the:main-1}
    Any algorithm $\cA$ that is $(\eps_1,\eps_2,\delta,\beta)_{n,d}$ differentially-private for  $\beta^2 \leq n$ 
    must satisfy $$\delta \geq\frac{(d-\beta)(n-d-\beta)-\beta^2}{d(n-d)}-2\eps_1-(e^{\eps_2}-1).$$
    Furthermore, there exists an  $(\eps_1,\eps_2,\delta,\beta)_{n,d}$  differentially-private  algorithm for $\eps_1=0$, $\eps_2=0$, and $$\delta = \left(1-\frac{\beta}{d}\right)\left(1-\frac{\beta}{n-d}\right)= \frac{(d-\beta)(n-d-\beta)}{d(n-d)}.$$
\ET
The proof of Theorem~\ref{the:main-1} appears in Section~\ref{sec:proof:Theorem1}. The Algorithm that satisfies the second condition of Theorem~\ref{the:main-1} is fairly simple. In particular,  given $e$, the algorithm returns a subset $e'$ chosen uniformly at random from the set system $\{e' \in E \mid d(e,e') = \beta\}$. We refer to this algorithm as Algorithm~${\cA}_1$.


The problem DPSR-I, 
is an extension of problem DPSR where the algorithm $\cA(e)$ is of a specific form in which the noise added to the set $e$ does not depend on $e$. Formally, in Problem DPSR-I, the algorithm  $\cA(e)$ can be written as 
\begin{equation}
\cA(e)=e\cup B,
\end{equation}
where $B\subseteq V$ is a random subset of $V$, which is chosen statistically independently of $e$. We refer to any algorithm that satisfies this requirement as a DPSR-I algorithm. 

Our second result is Theorem~\ref{the:main2} that characterizes the privacy-accuracy trade-off for Problem~DPSR-I.

In the theorem below, we consider DPSR-I algorithms that use subsets $B$ of size at most $\beta$. 
For such algorithms $\cA$ it follows that $d(\cA(e),e)$ is always (with probability 1) at most $\beta$ and thus the accuracy requirement of  Definition~\ref{def:gen_gt_private_f} always holds (i.e., corresponding to  $\eps_1=0$). Similarly, it holds, if $d+\beta<n$, that any DPSR-I algorithm for which  $d(\cA(e),e)$ is always at most $\beta$ must use subsets $B$ which are at most of size $\beta$. 
\BT
\label{the:main2}
Any DPSR-I algorithm $\cA$ that is $(\eps_1=0,\eps_2,\delta,\beta)_{n,d}$ differentially-private for $d+\beta<n$ must have $$\delta \geq 1- \frac{\beta}{n-d}-(e^{\eps_2}-1).$$ Moreover, there exists an $(\eps_1,\eps_2,\delta,\beta)_{n,d}$ differentially-private  DPSR-I Algorithm ${\cA}$ for $\eps_1=0$, $\eps_2=0$, $2d\beta^2<n$, and $\delta = 1- \frac{\beta-1}{n-d}$.
\ET

The proof of Theorem~\ref{the:main2} appears in Section~\ref{sec:thm2}. The algorithm that satisfies the conditions of Theorem~\ref{the:main2} is simple as well. Given $e$, the algorithm returns a subset $e'$ that is the union of $e$ and a subset $B \subset V$ of size  $\beta$ chosen uniformly at random from all subsets of $V$ of size  $\beta$. We refer to this algorithm as Algorithm ${\cA}_2$. 

In Section~\ref{sec:group_testing}, we define differential-private group-testing and use Theorems \ref{the:main-1} and \ref{the:main2} to derive our results in this context.


\isitremove{
\section{Proof of Theorem~\ref{the:main-1}}\label{sec:proof:Theorem1}

\subsection{Proof of Theorem~\ref{the:main-1}: Upper bound on $\delta$}

\begin{lem}[Upper bound of Theorem~\ref{the:main-1}]\label{lemma:1}
Algorithm $\cA_1$  satisfies Definition~\ref{def:gen_gt_private_f} with parameters $(\eps_1=0,\eps_2=0,\delta,\beta)$ for
    $$
\delta \leq \frac{(d-\beta)(n-d-\beta)}{d(n-d)}.
    $$
\end{lem}

\begin{prf}
Condition (i) in Definition~\ref{def:gen_gt_private_f} immediately follows by the definition of $\cA_1$. 
We now address condition (ii) in  Definition~\ref{def:gen_gt_private_f}.

For a subset $e_j$ in $N_1(e_i)$, we define a family of sets $E_{j,\alpha}^i$ for $1\le\alpha\le\beta$ as follows:

\begin{equation}\label{eq:1}
E_{j,\alpha}^i=\{e_l\in E| d(e_i,e_l)\le\alpha\ {\rm and}\ d(e_j,e_l)>\alpha    \}. 
\end{equation}

For any $e_1,e_2\in E$ with $d(e_1,e_2)=1$ 
and any 
$S\subset 2^V$ s.t for any $s\in S$, $|s|\ge d$ we show that $\Pr_A[{\cal A}_1(e_1)\in S]\le \Pr_A[{\cal A}_1(e_2)\in S]+\delta$ (recall that $\eps_2=0$).

First, notice that, by the definition of $\cA_1$, for $s\in S$ with $
|s|>d$ it holds that $\Pr_A[{\cal A}_1(e)=s]
=0$ for any $e\in E$. Therefore, for any $e\in E$ it holds that $$\Pr_A[{\cal A}_1(e)\in S]=\Pr_A[{\cal A}_1(e)\in S\cap E].$$ Thus we get:

\begin{align*}
& \Pr_A[{\cal A}_1(e_1)\in S]- \Pr_A[{\cal A}_1(e_2)\in S]\\
& = \Pr_A[{\cal A}_1(e_1)\in S\cap E]- \Pr_A[{\cal A}_1(e_2)\in S\cap E]\\
& = \sum_{e'\in S\cap E}{\Pr_A[{\cal A}_1(e_1)=e']}- \sum_{e'\in S\cap E}{\Pr_A[{\cal A}_1(e_2)=e']} \\
& \leq \sum_{e'\in E^1_{2,\beta}}{\Pr_A[{\cal A}(e_1)=e']}
= \frac{|E_{2,\beta}^1|}{|\bigcup_{\alpha=1}^{\beta}N_{\alpha}(e_1)\cup\{e_1\}|}\\
& = \frac{{d-1 \choose d-\beta-1}{n-(d+1)\choose \beta}}{\sum_{\alpha=1}^{\beta}{{d \choose \alpha}\cdot{n-d \choose \alpha}}+1}
\le \frac{{d-1 \choose d-\beta-1}{n-(d+1)\choose \beta}}{{d \choose \beta}\cdot{n-d \choose \beta}}\\
& = \frac{{d-1 \choose d-\beta-1}{n-(d+1)\choose \beta}}{{d \choose d-\beta}\cdot{n-d \choose \beta}}
= \frac{(d-\beta)(n-d-\beta)}{d(n-d)}
\end{align*}

Therefore, for $\delta=\frac{(d-\beta)(n-d-\beta)}{d(n-d)}$ (and $\eps_2=0$), we get that $\Pr_A[{\cal A}_1(e_1)\in S]\le e^{\eps_2}\cdot \Pr_A[{\cal A}_1(e_2)\in S]+\delta$.
\end{prf}

\subsection{Proof of Theorem~\ref{the:main-1}: Lower bound on $\delta$}
\label{sec:lower1}

We first prove the lower bound of Theorem~\ref{the:main-1} under the restriction that $\cA$ is $(\eps_1=0,\eps_2=0,\delta,\beta)_{n,d}$ differentially private, and always returns a subset of size $d$.
We later remove these restrictions.

\begin{lem}\label{lemma:lb0}
Let $\beta^2 \leq n$.
If $\cA$ is $(\eps_1=0,\eps_2=0,\delta,\beta)_{n,d}$ differentially private according to Definition~\ref{def:gen_gt_private_f}, and always returns a subset of size $d$, then
    $$
\delta \geq \frac{(d-\beta)(n-d-\beta)-\beta^2}{d(n-d)}.
    $$
\end{lem}

\begin{prf}
For a subset $e_j$ in $N_1(e_i)$, we define a family of sets $E_{j,\alpha}^i$ for $1\le\alpha\le\beta$ as follows:

\begin{equation}\label{eq:1a}
E_{j,\alpha}^i=\{e_l\in E| d(e_i,e_l)\le\alpha\ {\rm and}\ d(e_j,e_l)>\alpha    \}. 
\end{equation}

Note that for any $e_l\in E_{j,\alpha}^i$ it holds that $d(e_i,e_l)=\alpha$ and $d(e_j,e_l)=\alpha+1$.


For each $e_i$ and $e_j$, we define

$$
P_{ij}=Pr[{\cal A}(e_i)=e_j].
$$

Note that $\forall e_i\in E$ it holds that 

$$
\sum_{e_j\in E}P_{ij}=1.
$$




Then, by applying the second condition of  differential privacy to the set $E'=\{e_i\}\cup \bigcup_{1\le\alpha\le\beta} E_{j,\alpha}^i$ we get:

\begin{equation}\label{eq:2B}
P_{ii}+\sum_{\alpha=1}^{\beta}\sum_{e_l\in E_{j,\alpha}^i}P_{il}\leq P_{ji}+ \sum_{\alpha=1}^{\beta-1}\sum_{e_l\in E_{j,\alpha}^i}P_{jl}+\delta.
\end{equation}

Now, let us sum Equation~\eqref{eq:2B} over all $e_j\in N_1(e_i)$

We get 

\begin{align}
\label{eq:3}
d(n&-d)P_{ii}+\sum_{\alpha=1}^{\beta}\sum_{e_j\in N_1(e_i)}\sum_{e_l\in  E_{j,\alpha}^i}P_{il}\\
& \leq \sum_{e_j\in  N_1(e_i)}P_{ji}+ \sum_{\alpha=1}^{\beta-1}\sum_{e_j\in N_1(e_i)}\sum_{e_l\in E_{j,\alpha}^i}P_{jl}+d(n-d)\delta.\nonumber
\end{align}

We prove that 
there are exactly $(d-\alpha)(n-d-\alpha)$ subsets $e_j\in N_1(e_i)$ that satisfy $e_l\in E_{j,\alpha}^i$, for any $1\le\alpha\le\beta$. 
First, note that $e_j$ must include exactly $d-1-\alpha$ elements of $e_i\cap e_l$:
\begin{itemize}
    \item if $e_j$ includes less than $d-1-\alpha$ elements of $e_i\cap e_l$ then $d(e_i,e_j)>1$ (this is true since $e_j\cap e_i=(e_j\cap e_i\cap e_l)\cup(e_j\cap(e_i\setminus e_l))$, so if $|e_j\cap e_i\cap e_l|< d-1-\alpha$ it holds that $|e_j\cap e_i|=|e_j\cap e_i\cap e_l|+|e_j\cap(e_i\setminus e_l)|<(d-1-\alpha)+\alpha=d-1$,
    \item if $e_j$ includes more than $d-1-\alpha$ elements of $e_i\cap e_l$ then $d(e_l,e_j)\le \alpha$.
\end{itemize}  In addition, $e_j$ includes exactly one element not in $e_i$. This element cannot be an element of $e_l$, otherwise $d(e_l,e_j)\le \alpha$. Therefore this element is in $V\setminus(e_i\cup e_l)$. 
The rest of the elements of $e_j$ are in $e_i\setminus e_l$. Since we need to choose $d-(d-\alpha-1)-1=\alpha$ elements, and $|e_i\setminus e_l|=\alpha$, $e_j$ includes all the elements of $e_i\setminus e_l$. 
Therefore there are ${d-\alpha\choose d-\alpha-1}\cdot{n-(d+\alpha)\choose 1}=(d-\alpha)(n-d-\alpha)$ subsets  $e_j\in N_1(e_i)$, that satisfy $e_l\in E_{j,\alpha}^i$.

Thus, each $P_{il}$ in the sum $\sum_{e_j\in N_1(e_i)}\sum_{e_l\in  E_{j,\alpha}^i}P_{il}$ on the LHS of (\ref{eq:3}) is counted exactly $(d-\alpha)(n-d-\alpha)$
 times. 
Therefore, 
\begin{equation}
\sum_{e_j\in N_1(e_i)}\sum_{e_l\in  E_{j,\alpha}^i}P_{il}=(d-\alpha)(n-d-\alpha)\sum_{e_l\in N_{\alpha}(e_i)}P_{il}.
\end{equation}

We use this fact 
to obtain:

\begin{align*}
         d(n-d)&P_{ii} + \sum_{\alpha=1}^{\beta}(d-\alpha)(n-d-\alpha)\sum_{e_l\in N_{\alpha}(e_i)}P_{il}\\
        & \le 
        \sum_{e_j\in  N_1(e_i)}P_{ji}+ \sum_{\alpha=1}^{\beta-1}\sum_{e_j\in N_1(e_i)}\sum_{e_l\in E_{j,\alpha}^i}P_{jl}+d(n-d)\delta.\\
\end{align*}
      


Since $\sum_{e_l\in N_{\beta}(e_i)} P_{il}=1-P_{ii}- \sum_{\alpha=1}^{\beta-1}\sum_{e_l\in N_{\alpha}(e_i)} P_{il}$ we get:

         \begin{align*}
         d(n&-d)P_{ii} + \sum_{\alpha=1}^{\beta-1}(d-\alpha)(n-d-\alpha)\sum_{e_l\in N_{\alpha}(e_i)}P_{il}\\
         & \ \ \ \ \ + (d-\beta)(n-d-\beta)(1-P_{ii}-\sum_{\alpha=1}^{\beta-1}\sum_{e_l\in N_{\alpha}(e_i)}P_{il})\\
         & \le  \sum_{e_j\in  N_1(e_i)}P_{ji}+ \sum_{\alpha=1}^{\beta-1}\sum_{e_j\in N_1(e_i)}\sum_{e_l\in E_{j,\alpha}^i}P_{jl}+d(n-d)\delta,
         \end{align*}
which implies,

{\small{
   \begin{align}
   \label{eq:5}
         (d-&\beta)(n-d-\beta)+(d(n-d)-(d-\beta)(n-d-\beta))P_{ii} \nonumber\\
         & \  + \sum_{\alpha=1}^{\beta-1}((d-\alpha)(n-d-\alpha)-(d-\beta)(n-d-\beta))\sum_{e_l\in N_{\alpha}(e_i)}P_{il}\nonumber\\
         & \le 
         \sum_{e_j\in  N_1(e_i)}P_{ji}+ \sum_{\alpha=1}^{\beta-1}\sum_{e_j\in N_1(e_i)}\sum_{e_l\in E_{j,\alpha}^i}P_{jl}+d(n-d)\delta.\nonumber\\
   \end{align}
}}

Finally, by summing (\ref{eq:5}) over all $e_i\in E$, we obtain:

{\small{
\begin{align*}\label{eq:6}
\mu&(d-\beta)(n-d-\beta)+(d(n-d)-(d-\beta)(n-d-\beta))\sum_{e_i\in E}P_{ii}\\
& \ +\sum_{\alpha=1}^{\beta-1}((d-\alpha)(n-d-\alpha)-(d-\beta)(n-d-\beta))\sum_{e_i\in E}\sum_{e_l\in N_{\alpha}(e_i)}P_{il}\\
& \leq 
\sum_{e_i\in E}\sum_{e_j\in N_1(e_i)} P_{ji} + \sum_{\alpha=1}^{\beta-1}\sum_{e_i\in E}\sum_{e_j\in N_1(e_i)}\sum_{e_l\in E_{j,\alpha}^i}P_{jl}+\mu d(n-d)\delta.
\end{align*}
}}
 
Since

$$
\sum_{e_i\in E}\sum_{e_j\in N_1(e_i)}P_{ji}=\sum_{e_j\in E}\sum_{e_i\in N_1(e_j)}P_{ji}
$$

we have:

{\small{
\begin{align*}
\mu&(d-\beta)(n-d-\beta)+(d(n-d)-(d-\beta)(n-d-\beta))\sum_{e_i\in E}P_{ii}\\
&\ + \sum_{\alpha=1}^{\beta-1}((d-\alpha)(n-d-\alpha)-(d-\beta)(n-d-\beta))\sum_{e_i\in E}\sum_{e_l\in N_{\alpha}(e_i)}P_{il}\\
& \leq \sum_{e_j\in E}\sum_{e_i\in N_1(e_j)}P_{ji}+ \sum_{\alpha=1}^{\beta-1}\sum_{e_j\in E}\sum_{e_i\in N_1(e_j)}\sum_{e_l\in E_{j,\alpha}^i}P_{jl}+\mu d(n-d)\delta.
\end{align*}
}} 

We now bound $\sum_{e_j\in E}\sum_{e_i\in N_1(e_j)}\sum_{e_l\in E_{j,\alpha}^i}P_{jl}$ for all $1\le \alpha\le \beta-1$.
For fixed $e_j$ and $e_l$ such that $d(e_j,e_l)=\alpha+1$  
we count the number of $e_i$'s with  $d(e_j,e_i)=1$ and $d(e_i,e_l)=\alpha$.
It holds that $|e_l\setminus e_j|=\alpha+1$, so $e_i$ contains all $d-(\alpha+1)$ elements of $e_j\cap e_l$, $\alpha$ elements of $e_j\setminus e_l$  and one element of $e_l\setminus e_j$. 
This is true since if $e_i$ contains less than $d-(\alpha+1)$ elements of $e_j\cap e_l$, then it must contain $e_j\setminus e_l$ implying $d(e_i,e_i) \geq \alpha+1$.
Therefore, the number of such $e_i$ is ${|e_j\setminus e_l|\choose \alpha}\cdot {|e_l\setminus e_j|\choose 1}=(\alpha+1)^2$.


This implies that $$\sum_{e_j\in E}\sum_{e_i\in N_1(e_j)}\sum_{e_l\in E_{j,\alpha}^i}P_{jl}=(\alpha+1)^2\sum_{e_j\in E}\sum_{e_l\in N_{\alpha+1}(e_j)}P_{jl}.$$
Thus we obtain:

{\small{
\begin{flalign}
 & \mu(d-\beta)(n-d-\beta)+(d(n-d)-(d-\beta)(n-d-\beta))\sum_{e_i\in E}P_{ii}&\\\nonumber
     & \ + \sum_{\alpha=1}^{\beta-1}((d-\alpha)(n-d-\alpha)-(d-\beta)(n-d-\beta))\sum_{e_i\in E}\sum_{e_l\in N_{\alpha}(e_i)}P_{il}&\\  
     & \leq \sum_{e_j\in E}\sum_{e_i\in N_1(e_j)}P_{ji}+ \sum_{\alpha=1}^{\beta-1}(\alpha+1)^2\sum_{e_j\in E}\sum_{e_l\in N_{\alpha+1}(e_j)}P_{jl}+\mu d(n-d)\delta. \label{eq:beta10}
\end{flalign}
}}

which, for $\beta\ge 2$, implies that

{\small{
\begin{align}
\mu&(d-\beta)(n-d-\beta)\nonumber\\
& \ +(d(n-d)-(d-\beta)(n-d-\beta))\sum_{e_i\in E}P_{ii}\nonumber\\
& \ +((d-1)(n-d-1)-(d-\beta)(n-d-\beta)-1)\sum_{e_i\in E}\sum_{e_l\in N_{1}(e_i)}P_{il} \nonumber\\
& \ +\sum_{\alpha=2}^{\beta-1}((d-\alpha)(n-d-\alpha)-(d-\beta)(n-d-\beta))\sum_{e_i\in E}\sum_{e_l\in N_{\alpha}(e_i)}P_{il} \nonumber\\
& \ -\sum_{\alpha=2}^{\beta-1}{\alpha^2}\sum_{e_i\in E}\sum_{e_l\in N_{\alpha}(e_i)}P_{il}\nonumber\\ \label{eq:beta6}
& \leq 
\beta^2\sum_{e_j\in E}\sum_{e_l\in N_{\beta}(e_j)}P_{jl}+
\mu d(n-d)\delta.
\end{align}
}}
Since 
for $\sqrt{n} \geq \beta\ge 2$ and all $1\le\alpha\le\beta-1$, 
\begin{align*}
(d&-\alpha)(n-d-\alpha)-(d-\beta)(n-d-\beta)-\alpha^2)\\
&\ge
(d-(\beta-1))(n-d-(\beta-1))\\
& \ \ \ \ \ \ \ \ \ -(d-\beta)(n-d-\beta)-(\beta-1)^2\\
& =n-\beta^2\ge 0,
\end{align*}
we conclude that in the LHS of \eqref{eq:beta6} the coefficients of 
$\sum_{e_i\in E}P_{ii}$,
$\sum_{e_i\in E}\sum_{e_l\in N_{1}(e_i)}P_{il}$, and
$\sum_{e_i\in E}\sum_{e_l\in N_{\alpha}(e_i)}P_{il}$ are all greater than 0.
Moreover, as $\forall e_j\in E$,  
$\sum_{e_l\in N_{\beta}(e_j)}P_{jl}\leq 1$ we have that
$\beta^2\sum_{e_j\in E}\sum_{e_l\in N_{\beta}(e_j)}P_{jl}\le \beta^2\mu$.
Thus,
\begin{align*}
\mu(d-&\beta)(n-d-\beta)
\leq 
\beta^2\mu+\mu d(n-d)\delta.
\end{align*}
implying
\begin{equation*}
    \delta\geq\frac{(d-\beta)(n-d-\beta)-\beta^2}{d(n-d)}.
\end{equation*}

For $\beta=1$, by \eqref{eq:beta10} we have 
{\small{
 \begin{align*}
\mu&(d-1)(n-d-1)+(d(n-d)-(d-1)(n-d-1))\sum_{e_i\in E}P_{ii}\\
& \leq \sum_{e_j\in E}\sum_{e_i\in N_1(e_j)}P_{ji}+ \mu d(n-d)\delta.
\end{align*}
}}

Since 
$\forall e_j\in E$ it holds that 
$\sum_{e_l\in N_{1}(e_j)}P_{jl}\leq 1$,
we conclude as before that $\sum_{e_j\in E}\sum_{e_l\in N_{1}(e_j)}P_{jl}\le \mu$.
Also, $d(n-d)-(d-1)(n-d-1)\ge 0$, which implies
\begin{equation*}
\mu(d-1)(n-d-1)\leq \mu +\mu d(n-d)\delta,
\end{equation*}
so 
 $\delta\geq\frac{(d-1)(n-d-1)-1}{d(n-d)}$.
\end{prf}

We now extend the lower bound of Lemma~\ref{lemma:lb0} to hold for any $\eps_1,\eps_2$ and any algorithm $\cA$ that may return a subset of size different from $d$. 
We do this in two steps. The restriction of $\eps_1=\eps_2=0$ is first addressed in Lemma~\ref{lemma:lb} below.
Then the restriction on the size of $|\cA(e)|$ is addressed in Lemma~\ref{lem:size_d} appearing afterwards.

\begin{lem}\label{lemma:lb}
Let $\beta^2 \leq n$.
    If $\cA$ is $(\eps_1,\eps_2,\delta,\beta)_{n,d}$ differentially private according to Definition~\ref{def:gen_gt_private_f}, and always returns a subset of size $d$, then

     \begin{align*}
          \delta \geq \frac{(d-\beta)(n-d-\beta)-\beta^2}{d(n-d)} -2\eps_1 -(e^{\eps_2}-1)
    \end{align*}
    
     \end{lem}
\begin{prf}
Assume  $\cA$ 
as asserted in the lemma statement.
Let ${\cal A'}:E\rightarrow 2^V$ be a slight modification of $\cA$ defined as follows:
${\cal A'}(e)={\cal A}(e)$ if $d({\cal A}(e),e) \le \beta$, and ${\cal A'}(e)=e$ otherwise.
Notice that, just as $\cA$, the modified $\cA'$ always returns a subset of size $d$.
We now show that $\cA'$ is differentially private with parameters $(\eps'_1=0,\eps'_2=0,\delta'=\delta+2\eps_1+(e^{\eps_2}-1),\beta)$, implying by Lemma~\ref{lemma:lb0} that 
\[
\delta+2\eps_1+(e^{\eps_2}-1) = \delta' \geq \frac{(d-\beta)(n-d-\beta)-\beta^2}{d(n-d)},
\]
which concludes our proof.

First, notice that for any 
 $e\in E$, $\Pr[d({\cal A}'(e),e)\le \beta]= 1$, implying that $\cA'$ satisfies Definition~\ref{def:gen_gt_private_f}(i)  with $\eps'_1=0$.

For Definition~\ref{def:gen_gt_private_f}(ii), let $S \subseteq 2^V$.
Let $e_i,e_j\in E$ with  $d(e_i,e_j)=1$. 
By Definition~\ref{def:gen_gt_private_f}(ii) applied to $\cA$, we note that
\begin{align*}
\Pr[{\cal A}(e_i)\in S] & \le e^{\eps_2}\cdot\Pr[{\cal A}(e_j)\in S]+\delta \\
& = (e^{\eps_2}-1+1)\cdot\Pr[{\cal A}(e_j)\in S]+\delta\\
& = \Pr[{\cal A}(e_j)\in S]+\delta + (e^{\eps_2}-1)\Pr[{\cal A}(e_j)\in S] \\
& \le \Pr[{\cal A}(e_j)\in S]+\delta+(e^{\eps_2}-1).
\end{align*}


Now, let $S_i=\{e\in S \mid d(e_i,e)\le \beta\}|$ and $S_j=\{e\in S \mid d(e_j,e)\le \beta\}$.
From $\Pr[{\cal A}(e_i)\in S]\le \Pr[{\cal A}(e_j)\in S]+\delta+(e^{\eps_2}-1)$ we have:

 \begin{align*}
          \Pr[{\cA'}(e_i)\in S] 
         & = 
        \Pr[{\cA'}(e_i)\in S_i] +\Pr[{\cA'}(e_i)\in S\setminus S_i]  \\
         & =
         \Pr[{\cA'}(e_i)\in S_i]   \\
         &=
         \Pr[{\cA}(e_i)\in S_i] +\eps_1\\
         & \leq
         \Pr[{\cA}(e_i)\in S] +\eps_1\\
         & \leq
          \Pr[{\cA}(e_j)\in S] +  \eps_1+\delta+(e^{\eps_2}-1)\\
           &= 
    \Pr[{\cA}(e_j)\in S_j] + \Pr[{\cA}(e_j)\in S \setminus S_j] \\
          & \ \ \ \ \ \ \ \ \ \ \ \ \ \ \ \ \ \ \ \ \ \ \ \ \ \ \ \ \ \ \ \ +\eps_1+\delta+(e^{\eps_2}-1)\\
      & \leq
      \Pr[{\cA}(e_j)\in S_j] + \eps_1 +\eps_1+\delta+(e^{\eps_2}-1)\\
     & \leq
      \Pr[{\cA'}(e_j)\in S] + 2\eps_1+\delta+(e^{\eps_2}-1),
    \end{align*}
implying that $\cA'$ satisfies Definition~\ref{def:gen_gt_private_f}(ii) with parameters $\eps'_2=0$ and $\delta'=\delta+ 2\eps_1+(e^{\eps_2}-1)$.
All in all, $\cA'$ is differentially private with parameters $(\eps'_1=0,\eps'_2=0,\delta'=\delta+2\eps_1+(e^{\eps_2}-1),\beta)$ as asserted.
\end{prf}

We finally remove the restriction on the size of $\cA(e)$ 
to obtain the lower bound stated in  Theorem~\ref{the:main-1}.

\begin{lem}[Lower bound of Theorem~\ref{the:main-1}]
\label{lem:size_d}
Let $\beta^2 \leq n$.
If $\cA$ is $(\eps_1,\eps_2,\delta,\beta)_{n,d}$ differentially private according to Definition~\ref{def:gen_gt_private_f}, then
     \begin{align*}
          \delta \geq \frac{(d-\beta)(n-d-\beta)-\beta^2}{d(n-d)} -2\eps_1 -(e^{\eps_2}-1)
    \end{align*}
    
     \end{lem}

\begin{prf}
Assume  $\cA$ 
as asserted in the lemma statement.
$\cA$ does not necessarily return a subset of size $d$.
Let $\hat{{\cal A}}:V\rightarrow 2^V$ be a slight modification of $\cA$ defined as follows.
The modified algorithm $\hat{\cA}$ on input $e$ first runs $\cA$ on $e$, and then (if needed) modifies the output $\cA(e)$ to be of size exactly $d$.
Specifically, if $\bar{d} \triangleq|\cA(e)| >d$ then $\bar{d}-d$ elements are removed from $\cA(e)$ uniformly at random, if $\bar{d} < d$ then  $d-\bar{d}$ elements are added to $\cA(e)$ uniformly at random (from $V \setminus \cA(e)$), finally if $\bar{d}=d$ then $\hat{\cA}(e)=\cA(e)$.
We proceed to show that $\hat{\cA}(e)$ is $(\eps_1,\eps_2,\delta,\beta)_{n,d}$ differentially private, which in combination with the fact that $\hat{\cA}$ always returns a subset of size $d$, implies by Lemma~\ref{lemma:lb} that 
\[
\delta \geq \frac{(d-\beta)(n-d-\beta)-\beta^2}{d(n-d)}-2\eps_1-(e^{\eps_2}-1) ,
\]
which concludes our proof.

For Definition~\ref{def:gen_gt_private_f}(i), it suffices to show that for any $e\in E$, $\Pr[d(\hat{\cA}(e),e) \le \beta]\ge 1-\varepsilon_1$.
Towards that end, we show that $d(\hat{\cA}(e),e) \leq d({\cA}(e),e)$.
If $\bar{d}=d$, the assertion follows trivially.
If $\bar{d}>d$, then 
\begin{align*}
\bar{\beta} & \triangleq d(\cA(e),e)\\
& =
\max\{|{\cA}(e) \setminus e|,|e \setminus {\cA}(e)|\}\\
& = |{\cA}(e) \setminus e| = |e \setminus {\cA}(e)| + \bar{d}-d,
\end{align*}
so removing $\bar{d}-d$ elements from $\cA(e)$ will reduce  
$|{\cA}(e) \setminus e|$ and increase $|e \setminus {\cA}(e)|$
by at most $\bar{d}-d$. Implying that 
\begin{align*}
d(\hat{\cA}(e),e)& =\max\{|\hat{\cA}(e) \setminus e|,|e \setminus \hat{\cA}(e)|\} \\
& \leq \max\{|{\cA}(e) \setminus e|,|e \setminus {\cA}(e)|+\bar{d}-d\} \\
& = \bar{\beta} = d({\cA}(e),e).
\end{align*}
If $\bar{d}<d$, then 
\begin{align*}
\bar{\beta} & \triangleq d(\cA(e),e)\\
& =
\max\{|{\cA}(e) \setminus e|,|e \setminus {\cA}(e)|\}\\
& = |e \setminus {\cA}(e)| = |{\cA}(e) \setminus e| + d-\bar{d},
\end{align*}
so adding $d-\bar{d}$ elements from $V\setminus \cA(e)$ will reduce $|e \setminus {\cA}(e)|$ and increase  
$|{\cA}(e) \setminus e|$ 
by at most $d-\bar{d}$. Implying that 
\begin{align*}
d(\hat{\cA}(e),e)& =\max\{|\hat{\cA}(e) \setminus e|,|e \setminus \hat{\cA}(e)|\} \\
& \leq \max\{|{\cA}(e) \setminus e|+d-\bar{d},|e \setminus {\cA}(e)|\} \\
& = \bar{\beta} = d({\cA}(e),e).
\end{align*}

For Definition~\ref{def:gen_gt_private_f}(ii), denote by $\cD$ the function applied to $\cA(e)$ in the process of generating $\hat{\cA}(e)$ (that is, $\cD$ includes either the addition or removal of elements from $\cA(e)$ if needed).
We show that $\hat{\cA}$ satisfies Definition~\ref{def:gen_gt_private_f}(ii) with parameters $\eps_2$ and $\delta$.
     First, for any $e_i$ and $e_j$ with $d(e_i,e_j)=1$ and any $s \in 2^V$ we define
     $\Delta(s)=\Pr[\cA(e_i)=s]-e^{\eps_2}\Pr[\cA(e_j)=s]$
     and $S^+=\{s \mid \Delta(s) \geq 0\}$.
     Then, we note that Definition~\ref{def:gen_gt_private_f}(ii) implies that
     $$
     \Pr[\cA(e_i)\in S^+]-e^{\eps_2}\Pr[\cA(e_j)\in S^+] = \sum_{s \in S^+}\Delta(s) \leq \delta.
     $$
     
     Now, for such $e_i$ and $e_j$ as above,
     and any $S \subseteq 2^V$: 
     \begin{align*}
         \Pr[\hat{\cA}(e_i)& \in S] 
         = 
         \sum_{s \in 2^V}\Pr[\cA(e_i)=s]\Pr[\cD(s)\in S] \\
         & =
         \sum_{s \in 2^V}(e^{\eps_2}\Pr[\cA(e_j)=s]+\Delta(s))\Pr[\cD(s)\in S] \\
         &=
         e^{\eps_2}\sum_{s \in 2^V} \Pr[\cA(e_j)=s]\Pr[\cD(s)\in S] \\
         &\ \ \ \ \ \ \ \ \ \ \ \ \ \ \ \ \ \ + \sum_{s \in 2^V} \Delta(s)\Pr[\cD(s)\in S] \\
         & \leq
         e^{\eps_2}\Pr[\hat{\cA}(e_j)\in S]  + \sum_{s \in S^+} \Delta(s) \Pr[\cD(s)\in S] \\
         & \leq
        e^{\eps_2}\Pr[\hat{\cA}(e_j)\in S] + \sum_{s \in S^+} \Delta(s) \\
        & \leq 
        e^{\eps_2}\Pr[\hat{\cA}(e_j)\in S]  + \delta,
     \end{align*}
implying that $\hat{\cA}$ satisfies Definition~\ref{def:gen_gt_private_f}(ii) with parameters $\eps_2$ and $\delta$.
All in all, $\hat{\cA}$ is differentially private with parameters $(\eps_1,\eps_2,\delta,\beta)$ as asserted. \end{prf}
}

\isitremove{
\section{Proof of Theorem~\ref{the:main2}}
\label{sec:thm2}

\subsection{Theorem~\ref{the:main2}: Upper bound on $\delta$}
\begin{lem}[Upper bound of Theorem~\ref{the:main2}]\label{lemma:2}
If $2d\beta^2<n$, then Algorithm $\cA_2$  satisfies Definition~\ref{def:gen_gt_private_f} with parameters $(\eps_1=0,\eps_2=0,\delta,\beta)$ for
    $$
\delta \leq 1-\frac{\beta-1}{n-d}.
    $$
\end{lem}

\begin{prf}
Condition (i) in Definition~\ref{def:gen_gt_private_f} immediately follows by the definition of $\cA_2$. 
We now address condition (ii) in  Definition~\ref{def:gen_gt_private_f}.


Let $e_1,e_2\in E$ with $d(e_1,e_2)=1$.
For $i=1,2$ and $0\le \alpha\le \beta$, define

\begin{equation}
S_{\alpha}^i=\{s\subseteq V| |s|=d+\alpha\ {\rm and}\ e_i\subseteq s   \}. 
\end{equation}

In addition, for $i \ne j \in \{1,2\}$, let


\begin{equation}
S_{j,\alpha}^i=\{s\subseteq V| |s|=d+\alpha\ {\rm and}\ e_i\subseteq s\  {\rm and}\ e_j\not\subseteq s   \}. 
\end{equation}

 
For any $S \subseteq 2^V$, we now show that $\Pr[{\cal A}_2(e_1)\in S]\le \Pr[{\cal A}_2(e_2)\in S]+\delta$ (recall that $\eps_2=0$).
For $0\le \alpha\le\beta$, let $p_{\alpha}$ be the probability that, for a given $e$,  $\cA_2(e)$ is of size exactly $d+\alpha$.
Then 
\begin{equation}
p_{\alpha}=\frac{{d\choose \beta-\alpha}{n-d\choose \alpha}}{{n\choose \beta}}
\end{equation}

\begin{align*}
& \Pr[{\cal A}_2(e_1)\in S]- \Pr[{\cal A}_2(e_2)\in S]\\
& = \sum_{s\in S}{\Pr[{\cal A}_2(e_1)=s]}- \sum_{s\in S}{\Pr[{\cal A}_2(e_2)=s]} \\
& \leq \sum_{s\in \cup_{\alpha=0}^{\beta} S^1_{2,\alpha}}{\Pr[{\cal A}_2(e_1)=s]}\\ 
& = \sum_{\alpha=0}^{\beta}\sum_{s\in S^1_{2,\alpha}}{\Pr[{\cal A}_2(e_1)=s]}\\
& = \sum_{\alpha=0}^{\beta}\frac{|S_{2,\alpha}^1|}{|S_{\alpha}^1|}
\cdot p_{\alpha}\\
& = \sum_{\alpha=0}^{\beta}\frac{{n-(d+1)\choose \alpha}}{{n-d\choose \alpha}}\cdot p_{\alpha}\\
& = \sum_{\alpha=0}^{\beta}\left( 1-\frac{\alpha}{n-d}\right)
\cdot p_{\alpha}\\
& = 1-\frac{1}{n-d}\sum_{\alpha=0}^{\beta}\alpha
\cdot p_{\alpha}
\end{align*}

Now, for $2d\beta^2<n$,
$$
\sum_{\alpha=0}^{\beta}\alpha p_{\alpha} \geq \beta p_\beta  \geq \beta\left(1-\frac{d\beta}{n-\beta}\right) \geq \beta-1,
$$
implying that,
$$
1-\frac{1}{n-d}\sum_{\alpha=0}^{\beta}\alpha p_{\alpha} \leq
1-\frac{\beta-1}{n-d}.
$$
This concludes the proof.
\end{prf}

\subsection{Theorem~\ref{the:main2}: Lower bound on $\delta$}

We first prove the lower bound of Theorem~\ref{the:main2} under the restriction that $\cA$ is $(\eps_1=0,\eps_2=0,\delta,\beta)_{n,d}$ differentially private.
The restriction of $\eps_2 = 0$ is then removed.

 \begin{lem}\label{lemma:lb01}
Any DPSR-I algorithm $\cA$ that is $(\eps_1=0,\eps_2=0,\delta,\beta)_{n,d}$ differentially-private for $d+\beta<n$ must have $\delta \geq 1- \frac{\beta}{n-d}.$
\end{lem}



\begin{prf}
For each $B\subseteq V$ 
define $P_{B}$ to be the probability that ${\cal A}$ picks $B$. First, notice that since $\cA$ is differentially private for $\eps_1=0$, then for $|B|>\beta$ we have $P_B = 0$. Here, we use the fact that $d+\beta < n$.

For each $e_i\in E$ and each subset $s\subseteq V$ define 
$$
P_{s,i}=\sum_{s\setminus e_i\subseteq B\subseteq s}P_{B}.
$$

Namely, $P_{s,i}$ is the probability $\Pr[{\cal A}(e_i)=s]$.
Let $e_j$ be a subset in $N_1(e_i)$. We define the set $S_j^i$ as follows:

\begin{equation}
S_j^i=\{s\subseteq V|e_i\subseteq s, e_j\not\subseteq s ,d\le|s|\le d+\beta\}. 
\end{equation}


Then, by applying the second condition of differential privacy to the set $S_j^i$ we get for any $i,j$ that:

\begin{equation}\label{eq:2ff}
\sum_{s \in S_j^i}P_{s,i} = \sum_{s\subseteq V: e_i\subseteq s, e_j\not\subseteq s ,d\le|s|\le d+\beta}P_{s,i}\leq \delta.
\end{equation}

Now, let us sum \eqref{eq:2ff} over all $e_j\in N_1(e_i)$.
We get, for a given $i$, that

\begin{equation}\label{eq:3ff}
\sum_{e_j\in N_1(e_i)}\sum_{s\subseteq V:e_i\subseteq s, e_j\not\subseteq s ,d\le|s|\le d+\beta}P_{s,i}\leq d(n-d)\delta.
\end{equation}
or equivalently

\begin{equation}\label{eq:4ff}
\sum_{\alpha=d}^{d+\beta}\sum_{e_j\in N_1(e_i)}\sum_{s\subseteq V:e_i\subseteq s, e_j\not\subseteq s, |s|=\alpha}P_{s,i}\leq d(n-d)\delta.
\end{equation}

We note that for each 
$s\subseteq V,e_i\subseteq s, e_j\not\subseteq s$, of size $\alpha$, $d\le\alpha\le d+\beta$, there are exactly ${d\choose d-1}(n-\alpha)=d(n-\alpha)$ subsets  $e_j\in N_1(e_i)$, that satisfy $e_j\not\subseteq s$. 
Thus, for a given $\alpha$, each $P_{s,i}$ of a subset $s$  of size $\alpha$ for a subset $e_i$ on the left hand side of \eqref{eq:4ff} is counted exactly $d(n-\alpha)$ times. 

We use this fact to obtain:

\begin{equation}\label{eq:5ff}
\sum_{\alpha=d}^{d+\beta}d(n-\alpha)\sum_{s\subseteq V:e_i\subseteq s, |s|=\alpha}P_{s,i}\leq d(n-d)\delta.
\end{equation}


Using the fact that $\sum_{s\subseteq V:e_i\subseteq s,d\le|s|\le d+\beta}P_{s,i}=1$, 
we obtain:

\begin{align*}
 \sum_{\alpha=d}^{d+\beta}&d(n-\alpha) \sum_{s\subseteq V:e_i\subseteq s, |s|=\alpha}P_{s,i}\\
& \geq d(n-d-\beta)\sum_{\alpha=d}^{d+\beta}\sum_{s\subseteq V:e_i\subseteq s, |s|=\alpha}P_{s,i} \\
& = d(n-d-\beta)\sum_{s\subseteq V:e_i\subseteq s, d\le|s|\le d+\beta}P_{s,i}
=  d(n-d-\beta)
\end{align*}
which implies by Equation~\eqref{eq:5ff}

\begin{equation}\label{eq:4'''}
d(n-d-\beta)\leq d(n-d)\delta.
\end{equation}
We conclude that:

\begin{equation*}
    \delta\geq\frac{n-d-\beta}{n-d}=1-\frac{\beta}{n-d}.
\end{equation*}
\end{prf}
}


We now extend the lower bound of Lemma~\ref{lemma:lb01} to hold for $\eps_2>0$. Roughly speaking, this is done by a standard tradeoff between $\delta$ end $\eps_2$ in the privacy constraint of Definition~\ref{def:gen_gt_private_f}.

\begin{lem}[Lower bound of Theorem~\ref{the:main2}]\label{lemma:lb3}
Any DPSR-I algorithm $\cA$ that is $(\eps_1=0,\eps_2,\delta,\beta)_{n,d}$ differentially-private for $d+\beta<n$ must have $\delta \geq 1- \frac{\beta}{n-d}-(e^{\eps_2}-1).$ 
\end{lem}


\begin{prf}
As $\cA$ is $(\eps_1=0,\eps_2,\delta,\beta)_{n,d}$ differentially private, it holds by Definition~\ref{def:gen_gt_private_f}(ii) that for $S \subseteq 2^V$ and
$e_i,e_j\in E$ with  $d(e_i,e_j)=1$, 
\begin{align*}
\Pr[{\cal A}(e_i)\in S] & \le e^{\eps_2}\cdot\Pr[{\cal A}(e_j)\in S]+\delta \\
& = (e^{\eps_2}-1+1)\cdot\Pr[{\cal A}(e_j)\in S]+\delta\\
& = \Pr[{\cal A}(e_j)\in S]+\delta + (e^{\eps_2}-1)\Pr[{\cal A}(e_j)\in S] \\
& \le \Pr[{\cal A}(e_j)\in S]+\delta+(e^{\eps_2}-1).
\end{align*}
This implies that 
$\cA$ is also $(0,0,\delta+e^{\eps_2}-1,\beta)_{n,d}$ differentially private,
implying by Lemma~\ref{lemma:lb01} that 
$$\delta+(e^{\eps_2}-1) \geq 1-\frac{\beta}{n-d},$$
which concludes our proof.
\end{prf}


\section{Group Testing with Differential Privacy}\label{sec:group_testing}


Our framework for differentially private subset retrieval lends itself naturally to a procedure for private Group Testing. 
Group testing has a significant potential to guarantee privacy by avoiding individual tests and, instead, relying on pool testing to estimate the affected groups. In addition, the group testing procedure can take advantage of the inherent noise associated with pooled testing to ensure the privacy of individuals.

In the context of Group Testing, instead of a database and an agent, we consider two actors: (i) a {\em specimen collector} that collects specimens 
from $n$ individuals, $d$ of which are assumed to be contaminated, and (ii) a \emph{lab} that has the capability to test specimens (or a combination thereof) for infection. 
 %
 %
The specimen collector does not have the capability to test vials for contamination and thus cannot gain any knowledge about the contaminated individuals.  Instead, the specimen collector has the capability to mix the content of vials and also to add ``noise'' in the form of replacing any given vial with a contaminated or uncontaminated vial. We will take advantage of the specimen collector's capabilities to add noise to satisfy the privacy requirements. The {lab} 
is interested in identifying which individuals in the population are contaminated. Our goal in this section is to establish a group testing privacy framework that enables the lab to identify the approximate set of infected individuals while ensuring that the lab cannot determine the infection status of any single individual with certainty exceeding a predefined threshold. We assume that no collusion is taking place between the sample collector and the lab.


\subsection{Group Testing Model}

In the group testing set up, the randomized algorithm \mbox{${\cal A}:E\rightarrow 2^V$} is split into two parts: (i) the \emph{testing function} $\mathcal{T}:E\rightarrow \{0,1\}^L$ that maps $e\in E$ to an $L$ dimensional \emph{syndrome} vector $S\in \{0,1\}^L$; (ii) the \emph{decoding} procedure $\mathcal{D}:\{0,1\}^L\rightarrow 2^V$ that maps $S\in \{0,1\}^L$ to a subset $e'\in 2^V$. Thus, $\cA(e)=\mathcal{D}(\mathcal{T}(e))$.
 %
Furthermore, the testing function is restricted to the sequence of $L$ pooled tests characterized by subsets $T_1,T_2,\dots,T_L$; each $T_i\in 2^V$ represents a subset of individuals whose specimens are mixed together. 
The (noiseless) testing procedure is defined as follows:
\begin{equation}\label{eq:no_noise}
\mathcal{T}(e)[i]=\begin{cases}
        0, & \text{if } T_i\cap e=\emptyset\\
        1, & \text{otherwise.}\\
        \end{cases}     
\end{equation}
Unfortunately, the noiseless procedure is not strong enough to satisfy the differential privacy conditions. 
Thus, to ensure privacy, the addition of noise is instrumental to the process described above. 
Accordingly, we consider the following two types of noise that can be added to the group testing procedure. We show later in Section~\ref{sec:implementation} how, in practice,  the specimen collector can add noise to the individual and pooled specimens. 
\textbf{Noise added before pooling.} Here, we consider a testing procedure $\mathcal{T}(e)$ in which one first randomly contaminates the specimens before performing the pooling and testing procedure. In particular,   a random subset $B$ of $V$ is chosen in a manner that is statistically independent of $e$; where the individuals in $B$ are considered to be contaminated, regardless of their actual infection status. The pooled testing is then performed on the set $e \cup B$ and satisfies:

\begin{equation}
\label{eq:noiseGT}
\mathcal{T}(e)[i]=\begin{cases}
        0, & \text{if } T_i\cap (e \cup B)=\emptyset\\
        1, & \text{otherwise.}\\
        \end{cases}     
\end{equation}





\textbf{Noise added after pooling.} Here, the results after pooling may be  randomly overwritten by noise.
Consider the \emph{indicator function} $I:2^V\rightarrow\{0,1\}$ such that for any $s\subseteq V$, it holds that $I(s)=0$ if $s=\emptyset$ and $I(s)=1$ otherwise. The noisy pooled testing now satisfies for (possibly dependent) random variables $U_1,\dots,U_L$ over $\mathcal{U}=\{0,1,2\}$,

\begin{equation}
\label{eq:noiseGT2}
\mathcal{T}(e)[i]=\begin{cases}
        0 & \text{when } U_i=0\\
        1 & \text{when } U_i=1\\
        I(T_i\cap e) & \text{when } U_i=2\\
        \end{cases}    
\end{equation}


As discussed in Section~\ref{sec:implementation}, both noisy models for testing function $\mathcal{T}(e)$ can be implemented without any information about $e$. Moreover, it is not hard to verify that the testing function defined by \eqref{eq:noiseGT} (in which the noise is added before pooling), can be viewed as a special case of the testing function defined by \eqref{eq:noiseGT2}, (in which the noise is added after pooling). Indeed, this can be accomplished  by defining random variables $U_1,\dots,U_L$ such that $U_i=1$ if $T_i\cap B\neq \emptyset$ and $U_i=2$ otherwise.

Decoding function $\mathcal{D}$ 
maps the syndrome $\mathcal{T}(e)$ to a subset $e'\in E$. The testing and decoding functions are designed jointly, i.e., the decoder  
has information about the testing sets $T_1,T_2,\dots,T_L$ and the type of the testing function. However, the decoder does not know the realization of the {randomness} in $\mathcal{T}$

 The goal of the group-testing privacy-preserving framework is to identify a randomized
 testing procedure $\mathcal{T}$ 
 and a corresponding decoding procedure $\mathcal{D}$, that satisfies the following differential privacy condition. 

\begin{definition}
\label{def:gen_gt_private_new}
An $(\eps_1,\eps_2,\delta,\beta)_{(n,d)}$ differentially-private solution to the group-testing problem is  
 a randomized testing procedure 
 $\mathcal{T}:E\longrightarrow \{0,1\}^L$ and a decoding procedure  
 $D:\{0,1\}^L\longrightarrow 2^V$
 that satisfy the following conditions:
 \begin{itemize}
 \item[(i)] (accuracy) for any 
 $e\in E$, $$\Pr[d({\cal D}({\cal T}(e)),e) \le \beta]\ge 1-\varepsilon_1,$$
 \item[(ii)] (privacy) for any $e',e''\in E$ 
 with $d(e',e'')=1$ and any $W \subseteq \{0,1\}^L$, it holds that
$$\Pr[{\cal T}(e')\in W]\le e^{\eps_2}\cdot\Pr[{\cal T}(e'')\in W]+\delta.$$ 
 \end{itemize}
Here, the probabilities are taken over the randomness of the testing function
$\mathcal{T}$.
\end{definition}

Note that this definition imposes the privacy requirement on the testing function and that if a testing function satisfies this requirement, the lab 
will not be able to determine {whether a single individual} is infected (with a probability that exceeds a certain threshold).



\BT
\label{the:main3}
Any testing function $\mathcal{T}:E\rightarrow \{0,1\}^L$ {(including testing functions defined by}  \eqref{eq:no_noise}, \eqref{eq:noiseGT}, and \eqref{eq:noiseGT2})   
    and any decoding function 
    $\mathcal{D}:\{0,1\}^L\rightarrow 2^V$ that are $(\eps_1,\eps_2,\delta,\beta)_{n,d}$ differentially-private for  $\beta^2 \leq n$ according to Definition~\ref{def:gen_gt_private_new}  
    must satisfy 
    \begin{equation}\label{equ:lower_bound}
    \delta \geq\frac{(d-\beta)(n-d-\beta)-\beta^2}{d(n-d)}-2\eps_1-(e^{\eps_2}-1).
    \end{equation}
    Moreover, there exists an $(\eps_1,\eps_2,\delta,\beta)_{n,d}$ differentially-private  testing  function $\mathcal{T}:E\rightarrow \{0,1\}^L$ as in \eqref{eq:noiseGT} and \eqref{eq:noiseGT2} and a decoding function 
    $\mathcal{D}:\{0,1\}^L\rightarrow 2^V$ for $\eps_1=0$, $\eps_2=0$, $2d\beta^2<n$, and $\delta = 1- \frac{\beta-1}{n-d}$.
\ET

\isitremove{
 \begin{prf} (sketch.) 
%
Let $\mathcal{T}$, $\mathcal{D}$ be $ (\eps_1,\eps_2,\delta,\beta)_{(n,d)}$ differentially private  testing and decoding problems according to  Definition~\ref{def:gen_gt_private_new}.
It is easy to verify that ${\cal A}(e)=\mathcal{D}(\mathcal{T}(e))$ is $ (\eps_1,\eps_2,\delta,\beta)_{(n,d)}$ differentially private according to Definition~\ref{def:gen_gt_private_f}.
Indeed, it is immediate by our definitions that the accuracy condition (i) for ${\cal A}$ holds; whereas the privacy condition (ii)
roughly follows from the fact that $\mathcal{D}$ is independent of $e$.  A similar argument was used in the proof of Lemma~\ref{lem:size_d}. 
We conclude, by Theorem~\ref{the:main-1}, the asserted lower bound on $\delta$.

In Section~\ref{sec:thm2}, we show that algorithm  
${\cA}_2$
(defined at the end of Section~\ref{sec:model}) satisfies  Theorem~\ref{the:main2}.
 %
 %
Accordingly, to prove the asserted claim, one can use the testing function 
$\mathcal{T}$ of \eqref{eq:noiseGT} in which  (i) $B$ is chosen uniformly at random from all subsets of $V$ of size  $\beta$ and (ii) the collection of subsets $T_1,\dots,T_L$ enables the reconstruction of $e\cup B$. Since the testing functions defined in \eqref{eq:noiseGT2} is a generalization of that of \eqref{eq:noiseGT}, the condition holds for the testing functions defined by \eqref{eq:noiseGT2} as well.
\end{prf}
}


\subsection{Implementation at the specimen collector}\label{sec:implementation}




To implement the noisy group testing functions 
$\mathcal{T}$ 
defined by \eqref{eq:noiseGT} and \eqref{eq:noiseGT2}, a specimen collector performs the following. First, a contaminated specimen can be obtained by mixing all vials collected. Indeed, the mixture of all vials will be contaminated since  $d$ of them are assumed to be contaminated. Second, an uncontaminated specimen can be obtained by using a neutral uncontaminated substance. 



To implement testing function 
$\mathcal{T}$
defined in \eqref{eq:noiseGT}, the specimen collector randomly chooses a subset $B$ 
of  
individuals and replaces their vials with a contaminated specimen. Then, the specimen collector 
assembles the pooling tests, as specified by sets $T_1,\dots, T_L$, and transfers them to the lab. 

To implement the noisy group testing procedure specified by \eqref{eq:noiseGT}, the specimen collector 
will switch combined specimens in pool $i$ with uncontaminated specimens when $U_i=0$ 
and with contaminated specimens
when $U_i=1$. When $U_i=2$, the specimen collector will transfer the original mix of  pool $i$ to the lab.

\section{Conclusion and Future Research}
\label{sec:diss}


The paper addresses subset retrieval and group testing problems under differential privacy constraints. These problems arise in practical scenarios where it is essential to identify an (approximate) group of infected individuals while ensuring certain privacy guarantees regarding the infection status of each individual. Privacy guarantees are crucial for several reasons, including increasing the likelihood of individual participation in testing by ensuring that their personal testing outcomes remain confidential. 

The paper makes the following contributions. 
First, we present a differential privacy framework for the subset retrieval problem 
and establish tight lower and upper bounds on the performance of any generic algorithm. Our results characterize the trade-off between the achievable degree of accuracy of the group representation and the degree of privacy that can be guaranteed to each individual. Next, we define a differential privacy framework for the noisy group testing problem. 
We show that our results for subset retrieval carry over to group testing as well. To the best of our knowledge, this is the first paper that considers group testing under the differential-privacy framework.

As part of future work, we aim to extend our framework to scenarios where the set of potentially contaminated subsets is limited and characterized by a hypergraph \cite{9834789}, in contrast to the arbitrary set-system of potentially contaminated subsets considered in this paper. Additional intriguing directions for future research include differentially private property testing and community detection.







\end{document}